\documentclass[fleqn,usenatbib]{mnras}
\usepackage{newtxtext,newtxmath}
\usepackage[T1]{fontenc}
\usepackage{ae,aecompl}

\usepackage{graphicx}	
\usepackage{amsmath}	
\usepackage{amssymb}	
\usepackage[utf8x]{inputenc}	

\def\ltsima{$\; \buildrel < \over \sim \;$}
\def\simlt{\lower.5ex\hbox{\ltsima}}
\def\gtsima{$\; \buildrel > \over \sim \;$}
\def\simgt{\lower.5ex\hbox{\gtsima}}

\def\msun{{\rm M_{\odot}}}

\def\del#1{{}}
\def\ltsima{$\; \buildrel < \over \sim \;$}
\def\simlt{\lower.5ex\hbox{\ltsima}}
\def\gtsima{$\; \buildrel > \over \sim \;$}
\def\simgt{\lower.5ex\hbox{\gtsima}}

\title[Evolution of dwarf galaxy parameters]{Evolution of dwarf galaxy observable parameters}

\author[E. Ledinauskas \& K. Zubovas]{
Eimantas Ledinauskas$^{1, 2}$\thanks{E-mail: eimantasl@gmail.com} and
Kastytis Zubovas$^{1, 2}$
\\
$^{1}$Center for Physical Sciences and Technology, Saulėtekio av. 3, Vilnius LT-10257, Lithuania\\
$^{2}$Vilnius University Observatory, Saulėtekio av. 3, Vilnius LT-10257, Lithuania
}

\date{Accepted XXX. Received YYY; in original form ZZZ}

\pubyear{2018}

\begin{document}
\label{firstpage}
\pagerange{\pageref{firstpage}--\pageref{lastpage}}
\maketitle

\begin{abstract}
We present a semi-analytic model of isolated dwarf galaxy evolution
and use it to study the build-up of observed correlations between
dwarf galaxy properties. We analyse the evolution using models with
averaged and individual halo mass assembly histories in order to
determine the importance of stochasticity on the present-day
properties of dwarf galaxies. The model has a few free parameters, but
when these are calibrated using the halo mass - stellar mass and
stellar mass-metallicity relations, the results agree with other
observed dwarf galaxy properties remarkably well. Redshift evolution
shows that even isolated galaxies change significantly over the Hubble
time and that `fossil dwarf galaxies' with properties equivalent to
those of high-redshift analogues should be extremely rare, or
non-existent, in the local Universe. A break in most galaxy property
correlations develops over time, at a stellar mass $M_* \simeq 10^7 \,
\msun$. It is caused predominantly by the ionizing background
radiation and can therefore in principle be used to constrain the
properties of reionization.  
\end{abstract}




\section{Introduction}

Studies of galaxy evolution in cosmological context usually focus on
massive galaxies ($M_* > 10^{9} \, M_\odot$) instead of dwarf galaxies
because the latter are much harder to detect and study observationally
due to their low luminosities \citep[e.g.,][]{Volgelsberger2014MNRAS,
  Schaye2015MNRAS}. However, with increasing amount and quality of
data the interest in dwarf galaxies also increases. They are
scientifically interesting and important because their evolution
depends on cosmological structure formation on the smallest scales and
so in principle their observed properties could be used to test
cosmological models in this still weakly constrained regime. For
example, research on dwarf galaxies could help illuminate the nature
and properties of dark matter \citep{WDM_simulations}. However,
galaxies are highly non-linear systems and so a solid theoretical
understanding of their evolution is needed to answer such
questions. This is especially true for dwarf galaxies as their
evolution is highly sensitive to complex baryonic matter processes due
to their shallow gravitational potentials
\citep[e.g.,][]{Pontzen2012MNRAS, Brooks2013ApJ, Sawala2015MNRAS}.

In recent years, a lot of progress has been achieved in understanding
dwarf galaxy properties and evolution. Cosmological-scale
representative volume simulations now reach baryonic mass resolutions
$M_{\rm res} < 10^5 \msun$, while zoom-in simulations reach $M_{\rm
  res} \sim 30 \msun$ \citep{Wheeler2019MNRAS}, which allows
investigation of galaxies with virial masses in the range $10^7 \msun
< M_{\rm vir} < 10^{10} \msun$ \citep{Wang2015MNRAS, Munshi2017arXiv,
  Wheeler2019MNRAS}. These simulations reproduce stellar-halo mass
relations that are consistent with observational data
\citep{Munshi2017arXiv, Garrison-Kimmel2019MNRAS}. Star formation in
the smallest galaxies is truncated by reionization
\citep{Efstathiou1992MNRAS, Thoul1996ApJ, Dijkstra2004ApJ,
  Hoeft2006MNRAS, Dawoodbhoy2018MNRAS, Wheeler2019MNRAS}, but even
these galaxies should build up stellar masses $M_* \sim 10^3 \msun$
before being quenched \citep{Ricotti2005ApJ, Wheeler2015MNRAS,
  Munshi2017arXiv}, although this result is sensitive to the star
formation prescription used \citep{Munshi2019ApJ}. More massive dwarf
galaxies have bursty star formation histories (SFHs)
\citep{Wheeler2019MNRAS, Garrison-Kimmel2019MNRAS}, consistent with
observations, and are predominantly quenched by stellar feedback
\citep{Wang2015MNRAS}, although supernova feedback alone might not be
able to quench them completely \citep{Bermejo2018MNRAS,
  Smith2019MNRAS}. Differences in merger histories lead to a spread in
final galaxy parameter distributions, with more massive and larger
galaxies forming when the halo assembly history is dominated by minor,
rather than major, mergers \citep{Cloet2014MNRAS}.

Observational campaigns allow investigation of individual dwarf
galaxies down to the level of individual dense cores
\citep{Rubio2015Natur} as well as examination of their
populations. These campaigns focus on nearby galaxies,
i.e. satellites of the Milky Way \citep{Newton2018MNRAS, Munoz2018ApJ}
or Local Group members \citep{Drlica2015ApJ}, with the MADCASH survey
being an exception \citep{Carlin2016ApJ}. Another recent important
result is the discovery of seven galaxy groups composed purely of
dwarf galaxies \citep{Stierwalt2017NatAs}, confirming theoretical
predictions that primordial density fluctuations should be scale-free
and that in some cases, dwarf galaxy populations can evolve in a
similar way to their more massive counterparts.

Such observations reveal numerous statistical correlations between
various galaxy parameters. In the case of massive galaxies, these can
be described by simple power-laws throughout wide parameter intervals
\citep[e.g.,][]{Finlator2008MNRAS,Elbaz2011A&A}. Dwarf galaxies may
also follow these correlations; however, because they might be
significantly affected by processes which are not very important for
massive galaxies, there might be a deviation at low masses from simple
power-laws in these relations which might yet be detected by future
surveys. There is some evidence for this deviation in, for example,
the star-forming main sequence \citep{McGaugh2017ApJ} and the
mass-metallicity relation \citep{Blanc2019ApJ}. One process which
might create this deviation is heating by the intergalactic
radiation field \citep{Barkana1999ApJ}. During the epoch of
reionization the average temperature of intergalactic gas rapidly
increases up to $\sim 2 \times 10^4$ K \citep{reionization_temp},
which is above the virial temperature of halos with masses lower than
$\sim 10^{10} \, M_\odot$. Gas accretion in these halos therefore
should be shut-off and the star formation rate (SFR) should rapidly
decrease \citep{gnedin2000, Dawoodbhoy2018MNRAS, Ivkovich2019MNRAS}.

In this work we use a semi-analytic model of dwarf galaxy evolution to
study how, in the standard $\Lambda$CDM cosmology, the statistical
relations between various galaxy parameters at the low-mass end depend
on cosmological redshift, and how these relations are affected by the
stochasticity of the dwarf galaxy halo mass assembly. We calibrate
some free parameters in the model using the stellar mass - halo mass
relation and stellar mass-metallicity relation. Our model reproduces
the observed galaxy stellar mass - gas mass relation and the stellar
mass - SFR relation very well, and approximately reproduces the
observed gas metallicities. Redshift evolution of these relations
shows that even isolated galaxies change significantly and `fossil
dwarf galaxies' left over from the early Universe should be extremely
rare. We also predict a break in most galaxy correlations at stellar
mass $M_* \sim 10^7 \, \msun$. This is consistent with observations,
although data for isolated dwarf galaxies with $M_* < 10^7 \, \msun$
is scarce. Our results will be testable with near-future observations
and should help better interpret their results.

The paper is organized as follows. In section~\ref{sec:model_desc} we
describe the semi-analytical model, in section~\ref{sec:calib_params}
we decribe the calibration of the free parameters. The main results -
analysis of observable parameter evolution and importance of
stochastic mass assembly - are presented in
section~\ref{sec:evolution_params}. Finally, we discuss our results in
the context of cosmological structure formation, provide predictions
for forthcoming surveys, and analyse the various assumptions made in
the model in section~\ref{sec:discussion}. We conclude in
section~\ref{sec:conclusions}.

Throughout the paper, we assume a standard $\Lambda$CDM cosmology with
cosmological parameters $H_0 = 67.74 \, \mathrm{km} \, \mathrm{s}^{-1}
\mathrm{Mpc}^{-1}$, $\Omega_\mathrm{b} = 0.05$, $\Omega_\mathrm{m} =
0.31,$ and $\Omega_\Lambda = 0.69$ from \citet{Planck2015}.

\section{Description of the model}
\label{sec:model_desc}

The semi-analytic model used here is very similar to the one we used
in \citet[hereafter Paper I]{EZ2018}. We therefore present it only
briefly, focusing mainly on the changes made and refer the reader to
Paper I for more detailed explanations.

We assume that dwarf galaxies are composed of 3 main components: a
dark matter halo, a gas disk and a stellar disk. In the subsections
below we describe the equations which are used to track the evolution
of these components in time. We use two versions of our model. In the
first version, the mass growth of dark matter halos and the merger
rates are calculated by using the average fitting formulas from
cosmological N-body simulations (section \ref{sec:dmh_growth}). In the
second version we use individual dark matter halo merger trees
(section \ref{sec:merger_trees}). The first version is used to
calibrate the free parameters and study the average galaxy evolution
trends and their average parameters while the second version is used
to study the effects of different possible mass growth
histories. Hereafter we call the first model `average' and the
second `stochastic'.

\subsection{Dark matter halo mass growth}
\label{sec:dmh_growth} 

The formation of dark matter halos, their mass assembly histories
(MAHs) and merger rates have been studied in detail with N-body
simulations and there are various fitting formulas that approximate
these processes rather well. In the average model we use a fitting
formula for dark matter halo mass growth rate from
\citet{avg_mahs_from_nbody}:
\begin{equation}
	\frac{\langle \dot{M}_{\rm dm} \rangle}{\msun \, {\rm yr}^{-1}} = 46.1 \left( \frac{M_{\rm dm}}{10^{12} \, \msun} \right)^{1.1} \left(1 + 1.11 z \right) \sqrt{\Omega_{\rm m} \left(1 + z \right)^3 + \Omega_\Lambda},
\end{equation}
where $M_{\rm dm}$ is the dark matter halo mass. This equation is
appropriate for dark-matter-only simulations. The halo mass function
in simulations including baryons shows that the total abundance of
halos of a given mass is $\sim 25\%$ lower; this also means that the
mass of halos with a given abundance is $\sim 25\%$ lower
\citep{Sawala2015MNRAS,Munshi2019ApJ}. We tested the importance of
this effect by rerunning the average models with a dark matter halo
growth rate reduced by $25\%$ and found that this makes almost no
difference to our results.

Two processes grow the dark matter halo mass: smooth accretion from
intergalactic medium (IGM) and merging with other halos. The average
rate of mergers with halos of a given mass is calculated by using the
fitting formula from \citet{merger_rate_from_nbody}:
\begin{equation}
\label{eq:merger_rate}
		\frac{\mathrm{d}^2 N_{\rm merg}}{\mathrm{d} \omega \mathrm{d} x} = A M_{12}^\alpha x^b \exp \left( (\tilde{x} / x)^\gamma \right) \,,
\end{equation}
where $N_{\rm merg}$ is the average number of mergers for a single
halo, $\omega$ is the monotonic function of time which can be used as
a natural time coordinate in Press–Schechter theory \citep[the precise
  definition and approximate formula for it can be found
  in][]{w_formula}, $x$ is the ratio between the larger and smaller
halo masses, $A = 0.065$, $\alpha = 0.15$, $b = -0.3$,
$\tilde{x}=2.5$, $\gamma = 0.5$ are fitted free parameters, $M$ is the
sum of the masses of the two halos. This relation fits both pure
N-body simulations as well as results of the Illustris simulations
that include baryonic processes \citep{merger_rate_illustris}.

\subsection{Merger trees} \label{sec:merger_trees}

To get the merger trees of individual halos for the stochastic models,
we used the publicly available code {\sc Pinocchio} \citep{PINOCCHIO}
which is based on the Lagrangian perturbation theory. Like in Paper I,
we used a run with box size $d = 28$ Mpc, sampled with $800^3$
particles. We assumed that the minimal halo comprises 10 particles,
which is standard value for {\sc Pinocchio}. This translates to the
smallest halo mass $M_{\rm min} \approx 2 \times 10^7 \, M_\odot$
which is small enough to resolve the merger trees of halos with the
final mass $M_{\rm dm,0} \geq 10^9 \, M_\odot$. As a result, this is
the lowest present-day mass of halos for which we have stochastic
models.

Also like in Paper I we used only halos which are isolated throughout
their whole lifetime. We define a halo as isolated if it is farther
than $2 r_{\rm vir}$ from all other more massive halos throughout the
Hubble time, where $r_{\rm vir}$ is the virial radius of the more
massive halo:
\begin{equation}
	r_{\rm vir} = \left( \frac{3 M_{\rm dm}}{4 \pi \rho_{\rm crit} \Delta_{\rm vir}} \right)^{1/3} \,,
\end{equation}
where $\rho_{\rm crit}$ is the critical density of the Universe and
$\Delta_{\rm vir}$ is the overdensity of the collapsed and virialized
spherical top-hat density fluctuation, which we calculated by using
fitting formula from \citet{Delta_vir}.

After running the {\sc Pinocchio} simulation, we selected subsets of
halos with final masses $M_{\rm dm}\left(z=0\right) \simeq 10^9 \,
\msun$, $2.5\times10^9 \, \msun$, $5\times10^9 \, \msun$, $10^{10} \,
\msun$ and $5\times10^{10} \, \msun$. We chose a small mass interval
around each value that included $N_{\rm h} = 300 \pm 10$ halos,
leading to $1500$ stochastic model halos in total.

\subsection{Dark matter halos}

Properties of dark matter halos are handled identically to Paper
I. Their density is assumed to follow the Navarro-Frenk-White (NFW)
profile \citep{nfw}, truncated at the virial radius. The concentration
parameter, which is defined as the ratio between virial radius and NFW
scale radius, was calculated by using the model of
\citet{dm_halo_concentration}, which relates the concentration
parameter $c$ to the time $t_{0.04}$ at which the halo assembled 4\%
of the mass it has at time $t$:
\begin{equation}
	c(t) = 4 \left[ 1 + \left( \frac{t}{3.75 t_{0.04}} \right)^{8.4} \right]^{1/8}.
\end{equation}

We characterize the gravitational potential of the dark matter halo by
its maximum circular velocity, which for the NFW profile can be
calculated by
\begin{equation}
	v_{\rm max} \approx 0.465 r_{\rm s} \sqrt{4 \pi G \rho_0} \,
\end{equation}
where $\rho_0$ is the normalization density of NFW profile. 

\subsection{Gas accretion and cooling}\label{sec:gas_accretion}

Similarly to dark matter, the gas mass of a galaxy increases because
of accretion from the IGM and mergers with other galaxies. Like
in Paper I, we assume that before the epoch of reionization smooth
accretion of gas is proportional to smooth accretion of dark matter:
\begin{equation}
	\dot{M}_\mathrm{g,acc} = \langle f_\mathrm{b} \rangle \dot{M}_\mathrm{dm,acc} \,,
\end{equation}
where $\langle f_\mathrm{b} \rangle \approx 0.19$ is the ratio between
the average densities of baryonic and dark matter in the Universe.

We also use the same algorithm as in Paper I to calculate the effect
of heating by the photoionizing intergalactic radiation field during and
after the epoch of reionization. This algorithm was first proposed in
\citet{okamoto2008}. If the virial temperature of a halo is lower than
the average temperature of intergalactic gas at overdensities
$\Delta_{\rm vir}/3$, then smooth accretion of baryonic matter is shut
off. If the virial temperature is lower than the intergalactic gas
temperature even at overdensities higher than $\Delta_{\rm evp} =
10^6$, the gas of the galaxy is evaporated exponentially:
\begin{equation}
	\dot{M}_\mathrm{evp} = \frac{M_\mathrm{g}}{r_\mathrm{vir} / c_\mathrm{s} (\Delta_{\rm evp})} \,,
\end{equation}
where $M_{\rm g}$ is the mass of the gas disk and $c_\mathrm{s} (\Delta_\mathrm{evp})$ is the sound speed of intergalactic gas at overdensities higher than $10^6$.

We assume that the accreted gas quickly cools and forms a disk with an
exponential surface density profile:
\begin{equation}
	\Sigma(R) = \Sigma_0 \exp \left(-R/R_{\rm s} \right) \,.
\end{equation}
We assume that stars also form a disk with a similar exponential
profile but with a different scale length. To calculate the scale
lengths of the disks we use relations fitted to observations from
\citet{Kravtsov2013}:
\begin{equation}
	R_{\rm s,*} = 0.011 r_{\rm vir} \,,\quad
	R_{\rm s,g} = 0.029 r_{\rm vir} \,.
\end{equation}

\subsection{Star formation and feedback}

SFR is calculated identically to Paper I, by using a model presented
in \citet{KrumholzSFLaw}; we refer the reader to this paper for a
detailed overview of the model. It agrees with observations and
assumes that SFR is directly related to the molecular gas fraction
which depends on the gas surface density, the metallicity, the star
formation rate and the gravitational potential.

The newly formed stars strongly affect the surrounding gas by feedback
in the form of winds, radiation and supernova explosions. Feedback is
a very important ingredient in the evolution of low-mass galaxies; it
must be included in models for them to reconstruct the observed
luminosity function of galaxies \citep{semi_analytic_review}. However,
stellar feedback is still not completely understood because modeling
it requires very high spatial resolution and it incorporates a lot of
different physical phenomena. As was shown in
\citet{feedback_comparison}, different stellar feedback
implementations can lead to very different results.

Here we explicitly model only supernova feedback. Other feedback modes
on smaller scales are included implicitly in the star formation
efficiency parameter when calculating SFR. We model the feedback due
to supernovae very crudely, by using an energy conservation
argument. We assume that ejected gas mass due to a single supernova
can be calculated by a relation:
\begin{equation}\label{eq:mej}
	m_\mathrm{ej} = \frac{2 \epsilon_{\rm sn} E_{\rm sn}}{ v_{\rm max}^2} \,,
\end{equation}
where $E_{\rm sn} = 10^{51} \, \mathrm{erg}$ is the approximate
kinetic energy generated by a single supernova and $\epsilon_{\rm sn}$
is a coefficient which determines what fraction of that energy is used
to eject the gas, as opposed to being radiated away in shocks as the
supernova remnant expands. The value of this coefficient should depend
on the gravitational potential of the galaxy. We parameterize this
dependence as a power-law:
\begin{equation}
	\label{eq:stellar_fb_eff}
	\epsilon_{\rm sn} = \mathrm{max} \left[ k_{\rm sn} \left( \frac{v_{\rm max}}{30 \, \mathrm{km/s}} \right)^{\gamma_{\rm sn}} ,\, 1 \right] \,,
\end{equation}
where $k_{\rm sn}$ and $\gamma_{\rm sn}$ are free parameters which we
calibrate by fitting the observed relation between stellar and dark
matter halo masses (section \ref{sec:calib_params}).

In this model we assume that supernovae completely eject the gas from
a galaxy, while in reality at least some of the heated gas might cool
down and fall back to the disk. However, as will be shown in section
\ref{sec:calib_params}, this treatment is sufficient to reconstruct
the observed statistical relations. There are two reasons for
this. First of all, fitting the feedback efficiency
(eq. \ref{eq:stellar_fb_eff}) means that we essentially recover the
fraction of supernova energy used to drive the gas out of the galaxy,
rather than the fraction of energy used to accelerate the gas, some of
which then might fall back. Secondly, gas fallback is more important
in massive galaxies \citep{MacLow1999ApJ, Efstathiou2000MNRAS}, and
the difference due to not modelling it is negligible for the dwarf
galaxies which are our focus in this paper.

We assume that only stars with masses $8 \, M_\odot < m_{\rm star} <
25 \, M_\odot$ explode as supernovae as more massive ones at typical
low dwarf galaxy metallicities should directly collapse to black holes
without a strong explosion \citep{massive_stars_end, Fryer2012ApJ};
the exact cutoff mass is highly uncertain and model-dependent, but our
results are essentially insensitive to it. Using the
\citet{Kroupa2001} stellar initial mass function and adding the type
Ia supernovae rate from \citet{SNIA_rate} we arrive at the total
supernova rate of 1 SN per 100 $M_\odot$ stars formed.

\subsection{Chemical evolution} \label{sec:chem_evolution}

The enrichment of interstellar medium (ISM) by metals and the recycled
mass during stellar evolution are calculated by using stellar
population synthesis program SYGMA \citep{SYGMA}. The newly-formed
metals are added to the gas disk and later are ejected or incorporated
into stars proportionally to the gas mass participating in these
processes. Only the total masses of metals in the disk and stars are
followed. We use the common instantaneous recycling approximation.

Stellar feedback also ejects metals from galaxies altogether. Because
ISM enrichment largely happens during supernova explosions, the
metallicity of ejected gas should be higher than the average
metallicity of the gas disk. This effect should be most important in
galaxies with shallow gravitational potentials. To model it we follow
\citet{krumholz_bathtub} and use the result of
\citet{metal_ejection_fraction} that the fraction of newly-formed
metals that are ejected strongly depends on the mass of a galaxy and
can be roughly calculated by:
\begin{equation}
	f_{\rm ej} = k'_{\rm ej} \, e^{-M_{\rm dm} / M_{\rm ej}} \,,
\end{equation}
where $k'_{\rm ej}$ and $M_{\rm ej}$ are fitting parameters. This
fraction should depend not only on the mass of a galaxy but also on
the concentration parameter. Because of this we choose to parametrize
$f_{\rm ej}$ not with mass but with maximal circular velocity $v_{\rm
  max}$:
\begin{equation}
	\label{eq:metals_ejected}
	f_{\rm ej} = k_{\rm ej} \, e^{-v_{\rm max} / v_{\rm ej}} \,.
\end{equation}
The value of $k_{\rm ej}$ determines the fraction of ejected metals in
the smallest galaxies.  In absence of any known reliable estimates of
this value we again follow \citet{krumholz_bathtub} and assume that
$k_{\rm ej} = 0.9$ as it is known from observations that even the
smallest galaxies were enriched so at least a small fraction of metals
should not be ejected.  With significantly higher values of this
parameter (e.g., $k_{\rm ej} = 0.99$) the models predict clearly too
low metallicities for low-mass galaxies and a value significantly
closer to 0 also seems implausible as the lowest mass galaxies clearly
should lose a large fraction of newly-formed metals.  We did not
calibrate this value to keep the free parameter count to a minimum and
already with this value the model reconstructs realistic
metallicities. The value of $v_{\rm ej}$, however, is calibrated by
fitting the observed stellar mass-metallicity relation (section
\ref{sec:calib_params}).

\subsection{Metallicity of the intergalactic medium} \label{sec:igm}

As gas is ejected from galaxies, the intergalactic medium (IGM) is
also enriched with metals over time. This means that gas accreted by
the model galaxy at later times has a higher than primordial
metallicity. This effect is important for dwarf galaxies, since their
low metallicity is sensitive to small changes in the metal content of
accreted material. Here we model only a subset of all galaxies so it
is impossible to model the IGM enrichment directly from the
model. Instead we estimate the evolution of IGM metallicity by
assuming that the rate of enrichment is proportional to the average
SFR density of the Universe. Then the IGM metallicity at certain
redshift can be calculated by:
\begin{equation}
	\label{eq:Z_igm}
	Z_{\rm igm}(z) = Z_{\rm prim} + (Z_{\rm igm,0} - Z_{\rm prim}) \int_{z_0}^{z} f_{\rm igm}(z') \mathrm{d} z' \,,
\end{equation} 
where $Z_{\rm prim} = 2\times10^{-5}$ is the primordial metallicity,
$Z_{\rm igm,0}$ is the IGM metallicity at $z=0$ (a free parameter),
$z_0$ is some starting redshift, and $f_{\rm igm}(z)$ is the function
proportional to the average SFR density of the Universe $\psi (z)$ at
redshift $z$:
\begin{equation}
	f_{\rm igm}(z) = k_{\rm igm} \psi (z) \frac{\mathrm{d} t}{\mathrm{d} z}
\end{equation}
where $k_{\rm igm}$ is chosen so that $Z_{\rm igm} = Z_{\rm igm,0}$
would hold at $z=0$:
\begin{equation}
	k_{\rm igm} = \left( \int_{z_0}^{0} \psi (z) \frac{\mathrm{d} t}{\mathrm{d} z} \mathrm{d} z \right)^{-1} \,\, \rightarrow \,\, \int_{z_0}^{0} f_{\rm igm}(z) \mathrm{d} z = 1.
\end{equation}
We calculate the average SFR density of the Universe by using a
fitting formula from various observations \citep{cosmic_SFH}:
\begin{equation}
	\psi(z) = 0.015 \frac{(1+z)^{2.7}}{1 + [(1+z) / 2.9]^{5.6}} \quad \left[M_\odot \mathrm{yr}^{-1} \mathrm{Mpc}^{-3} \right] \,.
\end{equation}
For this calculation a final IGM metallicity $Z_{\rm igm,0}$ must be
chosen. This value is calibrated by fitting the observed stellar
mass-metallicity relation (section \ref{sec:calib_params}). The value
of the primordial metallicity $Z_{\rm prim}$ has little influence on
the results as long as it is significantly lower than $Z_{\rm igm,0}$.

\subsection{Mergers of galaxies}\label{sec:mergers}

We treat galaxy mergers very approximately: during a merger, the more
massive halo incorporates all the dark matter, gas and stars of the
less massive halo. In reality, of course, mergers induce dynamical
perturbations which might affect the structure of a galaxy and induce
starbursts. However, as tested and described in Paper I, inclusion of
these effects would not affect the results significantly: even
allowing for extreme starbursts in major mergers results in only a
very small fraction of galaxies changing their final properties and
star formation histories. Therefore, in order to avoid complicating
the model and introducing many more free parameters for little
advantage in realism, we choose not to model the effects of mergers
explicitly. Even with the idealised approach, SFR usually increases
during mergers because of an abrupt increase in gas mass and gas
density.

In the average model we do not employ merger trees but the amount of
infalling mass during mergers is still calculated. This infalling mass
is calculated by using eq. (\ref{eq:merger_rate}):
\begin{equation}
	\dot{M}_{\rm i, merg}(z) = \int\displaylimits_1^{\infty} M_{\rm i}(x, z) \frac{\mathrm{d}^2 N_{\rm merg}}{\mathrm{d} \omega \mathrm{d} x} \frac{\mathrm{d} \omega}{\mathrm{d} t} \mathrm{d} x
\end{equation}
where $M_{\rm i}(x,z)$ is the mass of stars, gas or dark matter in
halos which are $x$ times less massive than the halo under
consideration. This treatment requires information about the evolution
of all less massive galaxies in order to know the masses of their
components at the time of merging. We solved this by always
calculating models in order of ascending mass and using the calculated
evolution data of all less massive galaxies than the one under
consideration. This treatment leads to a self-consistent sample of
galaxies.

\section{Calibration of free parameters} \label{sec:calib_params}

\begin{figure}
	\includegraphics[width=\columnwidth]{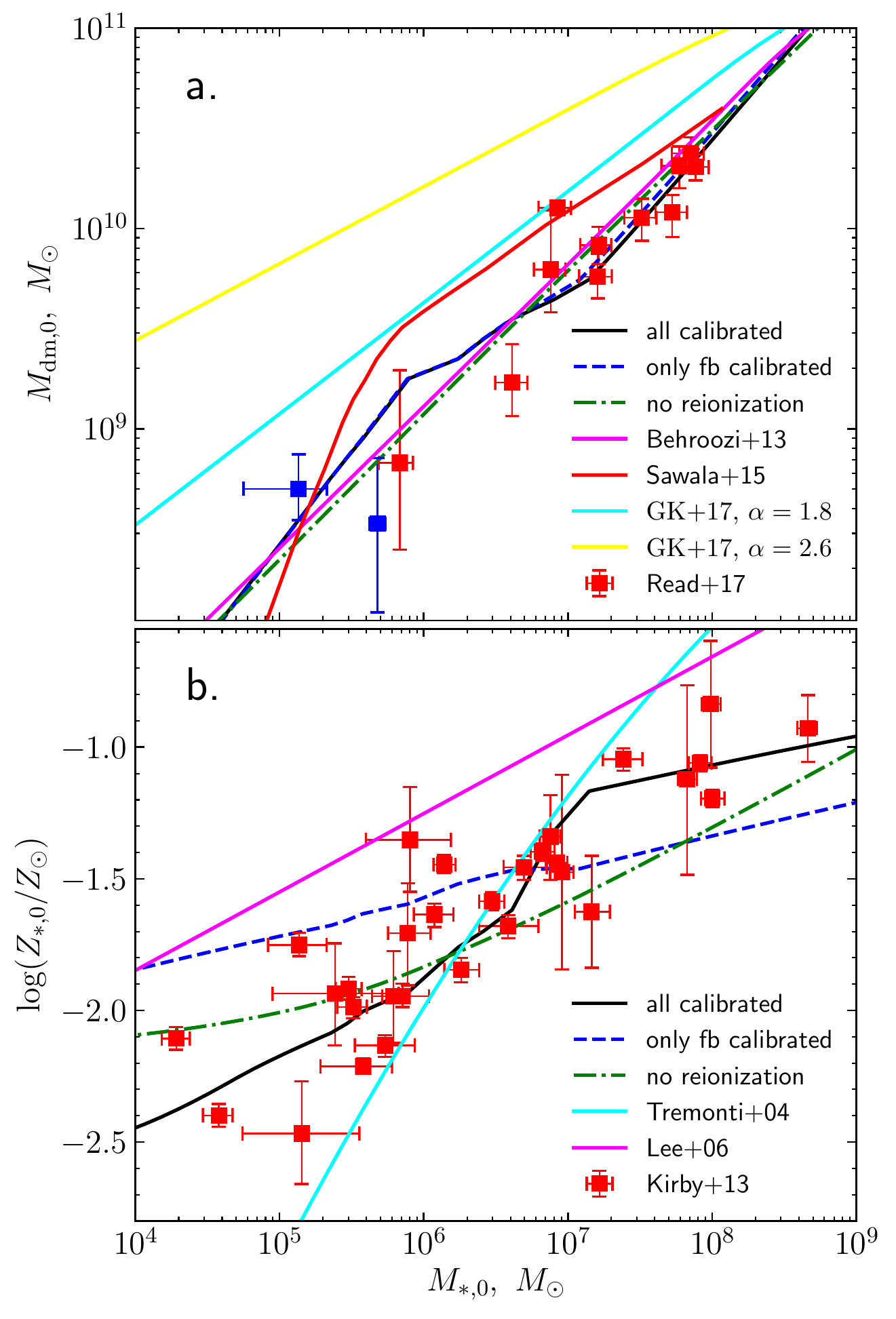}
    \caption{Relations between final stellar mass and dark matter halo
      mass (a) or stellar metallicity (b). Red and blue points
      represent data from observations \citep[in the top and bottom
        panels, respectively]{m_s_m_dm_relation,
        m_s_Z_s_relation}. Blue points denote non-isolated galaxies
      Carina and Leo T. Black line denotes the model with all four
      free parameters calibrated. Blue dashed line denotes the model
      with only $k_{\rm sn}$ and $\gamma_{\rm sn}$ calibrated. Green
      dash-dotted line denotes the model in which reionization is
      turned off and all 4 parameters recalibrated. In the top panel,
      magenta line shows the relation derived in
      \citet{Behroozi2013ApJ}, the red line a similar relation from
      \citep{Sawala2015MNRAS} and the cyan and yellow lines show
      relations from \citet{Garrison-Kimmel2017MNRAS}. In the bottom
      panel, the cyan line is from \citet{Tremonti2004ApJ} and the
      magenta line is from \citet{Lee2006ApJ}. See text for more
      details.}
    \label{fig:calib_params}
\end{figure}

Our model has four free parameters: $k_{\rm sn}$ and $\gamma_{\rm sn}$
describe the efficiency of mass ejection due to stellar feedback
(equation \ref{eq:stellar_fb_eff}), $v_{\rm ej}$ is the characteristic
maximal circular velocity below which galaxies lose the majority of
newly-formed metals (equation \ref{eq:metals_ejected}) and $Z_{\rm
  igm,0}$ is the present-day IGM metallicity (equation
\ref{eq:Z_igm}). To determine the realistic values of these parameters
we fit the relations between dark matter halo mass, mass-averaged
stellar metallicity\footnote{We assume $Z_\odot=0.0134$, as determined
  in \citet{asplund_solar_abund}} and stellar mass from the model to
the ones from observations.

The results are presented in fig. \ref{fig:calib_params}\footnote{To
  facilitate comparison, we converted the dark matter halo virial
  masses from our model into $M_{200}$, which is defined similarly as
  in this work, but $\Delta = 200$ is used instead of
  $\Delta_{\rm vir}$}. Red and blue points represent observational data
\citep[respectively in the top and bottom panels]{m_s_m_dm_relation,
  m_s_Z_s_relation}. Blue points denote non-isolated galaxies Carina
and Leo T. We fit the model parameters using orthogonal distance
regression\footnote{Open-source Python implementation available at
  https://docs.scipy.org/doc/scipy/reference/odr.html}. The black line
denotes a model when all four parameters were calibrated. The blue
line denotes a model when only $k_{\rm sn}$ and $\gamma_{\rm sn}$ are
calibrated; we set the ejected metal fraction $f_{\rm ej} = 0$
(eq. \ref{eq:metals_ejected}) and $Z_{\rm igm,0}=Z_{\rm prim}$. The
green line denotes a model in which the quenching of gas accretion and
photoevaporation due to reionization were turned off and all four
parameters were recalibrated. The relation between stellar and dark
matter masses depends very weakly on $Z_{\rm igm,0}$ and $v_{\rm ej}$,
as can be seen by comparing black and blue lines in
fig. \ref{fig:calib_params}, panel a. Because of this the stellar
feedback parameters $k_{\rm sn}$ and $\gamma_{\rm sn}$ can be fitted
independently from $Z_{\rm igm,0}$ and $v_{\rm ej}$ by using only the
relation between stellar and dark matter masses. After that, we fit
$Z_{\rm igm,0}$ and $v_{\rm ej}$ by using only the stellar
mass-metallicity relation. The determined parameter values are shown
in table \ref{tab:calib_params}. We also visually analyzed the grid of
parameters (intervals are shown in table \ref{tab:calib_params}) and
found that there are no degeneracies between parameter values and the
determined extremes are global.

For reference, we overplot equivalent relations from the literature in
the two panels of Figure \ref{fig:calib_params}. In the top panel, the
magenta line shows the low-mass extension of the relation $M_* \propto
M_{\rm h}^{1.4}$ from \citet{Behroozi2013ApJ}. This relation almost
perfectly coincides with our result for the no-reionization case and
is consistent with our other models, as expected. Similarly, the
relation derived by \citet[; red line]{Sawala2015MNRAS} is marginally
consistent with the data and with our models. On the other hand, the
two relations presented in \citet{Garrison-Kimmel2017MNRAS} - cyan
line for $M_* \propto M_{\rm h}^{1.8}$ and yellow line for $M_*
\propto M_{\rm h}^{2.6}$ - both lie significantly above our models and
the observational data, i.e. they predict much lower stellar masses at
a given halo mass. This difference is also expected, since
\citet{Garrison-Kimmel2017MNRAS} investigated satellite galaxies in
the Local Group, which are presumably significantly affected by tidal
stripping and other environmental quenching effects
\citep{Geha2012ApJ, Grossi2019IAUS}. In the bottom panel, the cyan
line shows the relation derived in \citet{Tremonti2004ApJ} for a large
sample of star-forming galaxies, while the magenta line shows the
relation from \citet{Lee2006ApJ}, derived for a small sample of nearby
dwarf irregulars. Both of these relations differ significantly from
our results and are much poorer matches to the observational data.

\begin{table}
	\centering
	\caption{The values of calibrated free parameters together
          with their grid intervals used and equation numbers in which
          they appear.}
	\label{tab:calib_params}
	\begin{tabular}{lccc}
		\hline\hline\\[-1em] 
		Parameter & Fitted value & Interval tested & Equation \\
		\hline\\[-1em]                    
		$k_{\rm sn}$ & 0.024 & [0, 1] & (\ref{eq:stellar_fb_eff}) \\ 
		$\gamma_{\rm sn}$ & 1.47 & [-3, 2] &  (\ref{eq:stellar_fb_eff}) \\  
		$v_{\rm ej}$ & 29 km/s & [10, 100] km/s &  (\ref{eq:metals_ejected}) \\   
		$Z_{\rm igm,0}$ & 0.0023 & [$2 \cdot 10^{-5}$, 0.01] &  (\ref{eq:Z_igm}) \\  
		\hline                      
	\end{tabular}
\end{table}

\begin{figure}
	\includegraphics[width=\columnwidth]{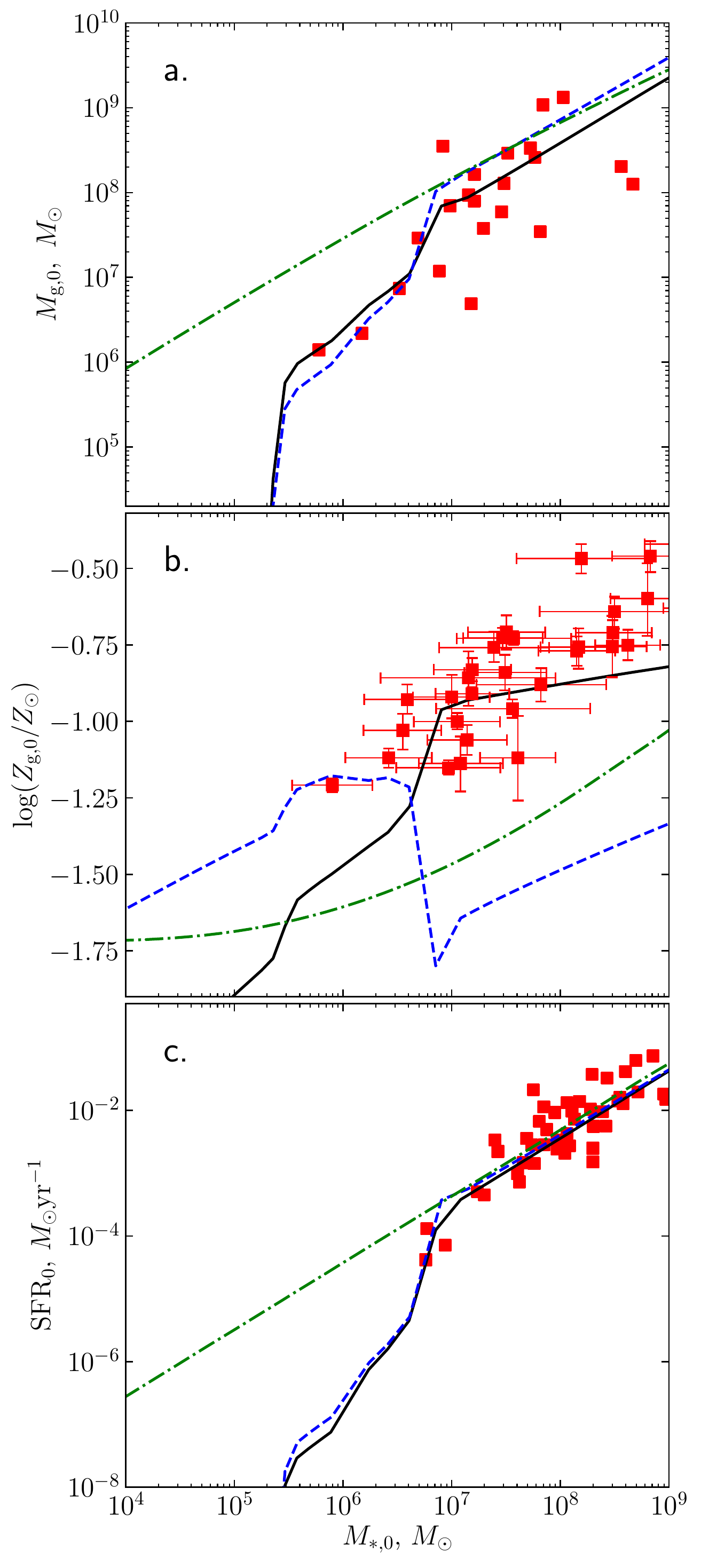}
    \caption{Relations between stellar mass and gas mass (a), gas metallicity (b), SFR (c) at $z=0$. Red points represent data from observations \citep{cuspvscore, Kirby2017, m_s_Z_g_relation, McGaugh2017ApJ}. Lines denote the same models as in fig \ref{fig:calib_params}.}
    \label{fig:check_calib}
\end{figure}

Figure \ref{fig:check_calib} shows the relations between gas mass, gas
metallicity, SFR and stellar mass at $z=0$. Once again, points are
observational data, from \citet{cuspvscore, Kirby2017,
  m_s_Z_g_relation, McGaugh2017ApJ} and the lines represent our model
results with different calibrations, as in Figure
\ref{fig:calib_params}. These three relations were not used for
parameter calibration and they are not simple consequences of the two
relations used for calibration.  Therefore, the good match between
observations and the fully calibrated model (black lines) in all these
relations strongly suggests that our model captures the salient
features of isolated dwarf galaxy evolution. Some discrepancy between
the best-fit model and observed gas metallicities most likely arises
due to our over-simplified treatment of enrichment, echoing the
results of earlier semi-analytic models \citep{Starkenburg2013MNRAS}.
Clear discrepancies between the model with reionization turned off and
the observations seen in panels a and b show that the modeling of the
effects of reionization is necessary in order to simultaneously
reconstruct all of the observed relations.

\begin{figure*}
	\resizebox{0.8\hsize}{!}
	{\includegraphics[]{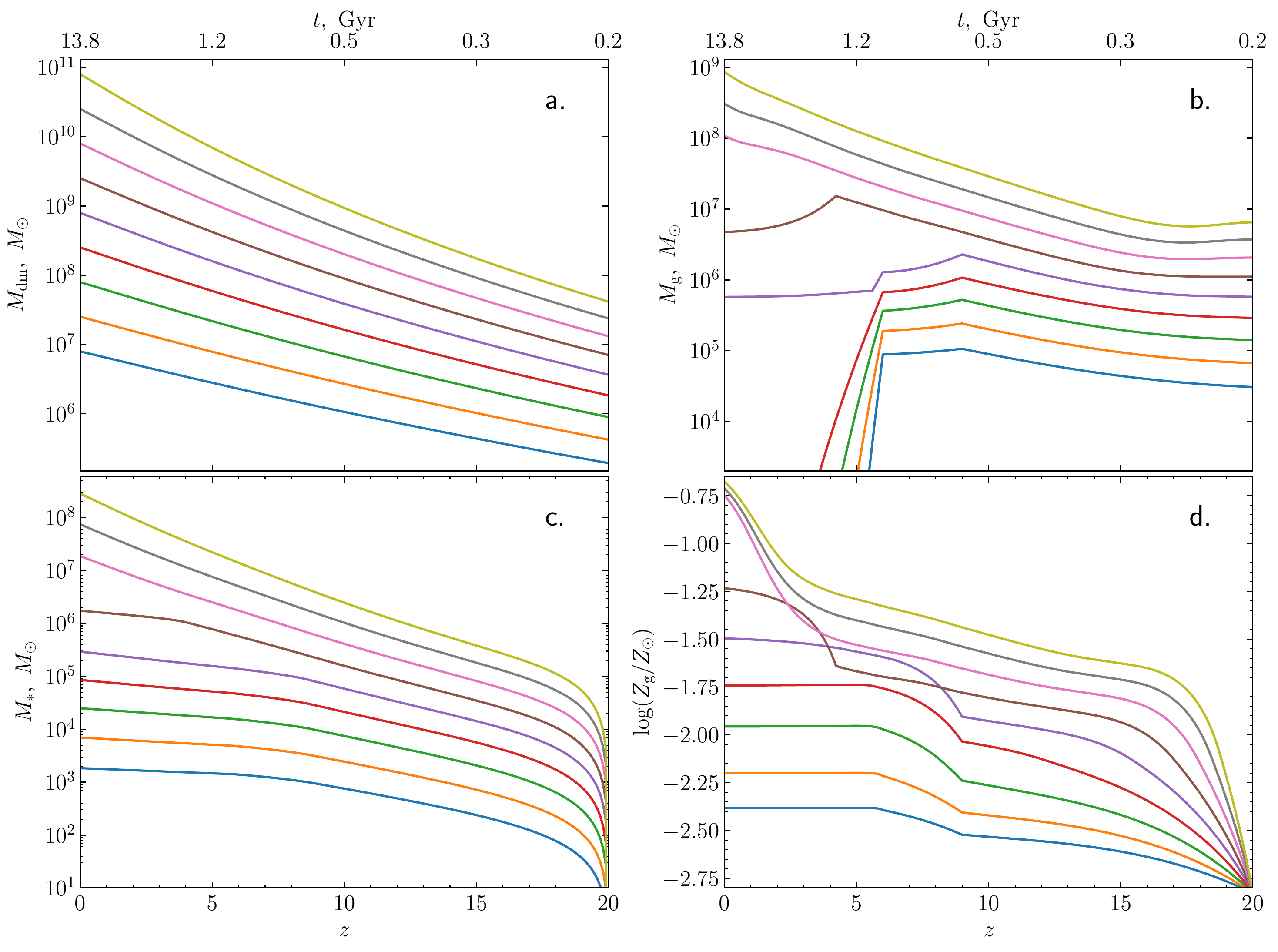}}
	\caption{Dependence of the dark matter halo mass, the gas mass, the stellar mass and the gas metallicity on redshift in the average model. Different colored lines denote models with different final dark matter halo masses.}
    \label{fig:avg_evolution}
\end{figure*}

All of the calibrated parameters have a clear physical interpretation
and therefore it is useful to discuss the implications of their
values. The values of $k_{\rm sn}$ and $\gamma_{\rm sn}$ determine the
fraction of supernova energy that is used to eject gas from a
galaxy. The amount of ejected gas per supernova scales as $m_{\rm ej}
\propto v_{\rm max}^{\gamma_{\rm sn}-2}$ (cf eq. \ref{eq:mej}).
Therefore, since $\gamma_{\rm sn} < 2$, the mass ejected per supernova
explosion decreases with increasing galaxy mass, as may be expeceted
due to the increasing gravitational potential. However, the efficiency
of ejection increases, since $\gamma_{\rm sn} > 0$. This might be
explained by the fact that in a larger galaxy there is a higher chance
for radiated supernova energy to be reabsorbed again and so on average
a higher fraction of supernova energy might be used to heat the
gas. The normalization of the efficiency, $k_{\rm sn} = 2.4\%$, is
similar to the analytically-derived $k\sim 3\%$ of the supernova
energy injected as kinetic energy into interstellar turbulence
\citep{Spitzer1990ARA&A}.

The present-day metallicity of the IGM is not known with certainty
because of difficulties in measuring it, but our calibrated value is
not implausible according to both observational and theoretical
estimates \citep{IGM_metallicity}. The calibrated value of $v_{\rm
  ej}$ also seems realistic because galaxies with lower maximal
circular velocities have virial temperatures close to or lower than
$10^4$ K which is a typical temperature of ionized interstellar
gas. Because the newly-formed metals are typically ejected into the
environments ionized due to stellar feedback, it seems plausible that
a major fraction of them are lost in galaxies with virial temperatures
lower than $10^4$ K.

\section{Evolution of observable parameters} \label{sec:evolution_params}

\subsection{Evolution of average models} \label{sec:average_evolution}

Having determined the values of the free parameters that give the best
results for modern-day dwarf galaxies, we now turn to the evolution of
model galaxy parameters over cosmic time. Figure \ref{fig:avg_evolution}
shows how the dark matter halo mass, the gas mass, the stellar mass
and the gas metallicity depend on redshift in the average model. The
nine lines in each panel correspond to models with different
present-day halo masses, from $8\times10^6 \, \msun$ to
$8\times10^{10} \, \msun$ in steps of $0.5$~dex.

The mass of the dark matter halo (Figure \ref{fig:avg_evolution}, panel
a) grows smoothly in all models because the effect of galaxy mergers
is smoothed over time and we model only isolated galaxies that do not
experience significant perturbations to their halos.

Evolution of the gas mass (panel b of Figure \ref{fig:avg_evolution})
shows several sudden breaks for galaxies with final halo masses below
$5\times10^9 \, \msun$ (fourth line from the top and lower). The
earliest break, at $z \approx 9$, occurs due to the start of the
reionization epoch. In the five lowest-mass models, smooth gas
accretion turns off (see section \ref{sec:gas_accretion}) and those
galaxies can then increase their gas masses only via mergers, at least
until their virial temperature grows above the threshold value. We
tested that our results are mostly insensitive to the precise redshift
of the start of reionization, by varying that parameter in the range
$6 < z_{\rm reion} < 15$. Reionization ends at $z=6$; this results in
a second break in gas mass, because from then on, galaxies with low
enough virial temperatures begin losing gas due to evaporation.
Evaporation can cease only if the virial temperature in the galaxy
grows large enough. This happens in the model with final halo mass
$8\times10^8 \, \msun$ (violet line, fifth from the top) at $z \simeq
5.5$, resulting in another sudden break in gas mass evolution.
Conversely, in a slightly more massive galaxy (brown line, fourth from
the top), virial temperature grows more slowly than $T_{\rm eq}\left(
\Delta_{\rm vir}/3 \right)$, eventually stopping the smooth accretion
of IGM gas at $z \simeq 4$, producing another break. Even in the three
most massive galaxies, where the virial temperature always remains
higher than $T_{\rm eq}\left( \Delta_{\rm vir}/3 \right)$, there are
small effects on gas mass due to reionization. They occur because a
fraction of the infalling gas comes from mergers with smaller galaxies
which are strongly affected by reionization.

\begin{figure}
	\includegraphics[width=\columnwidth]{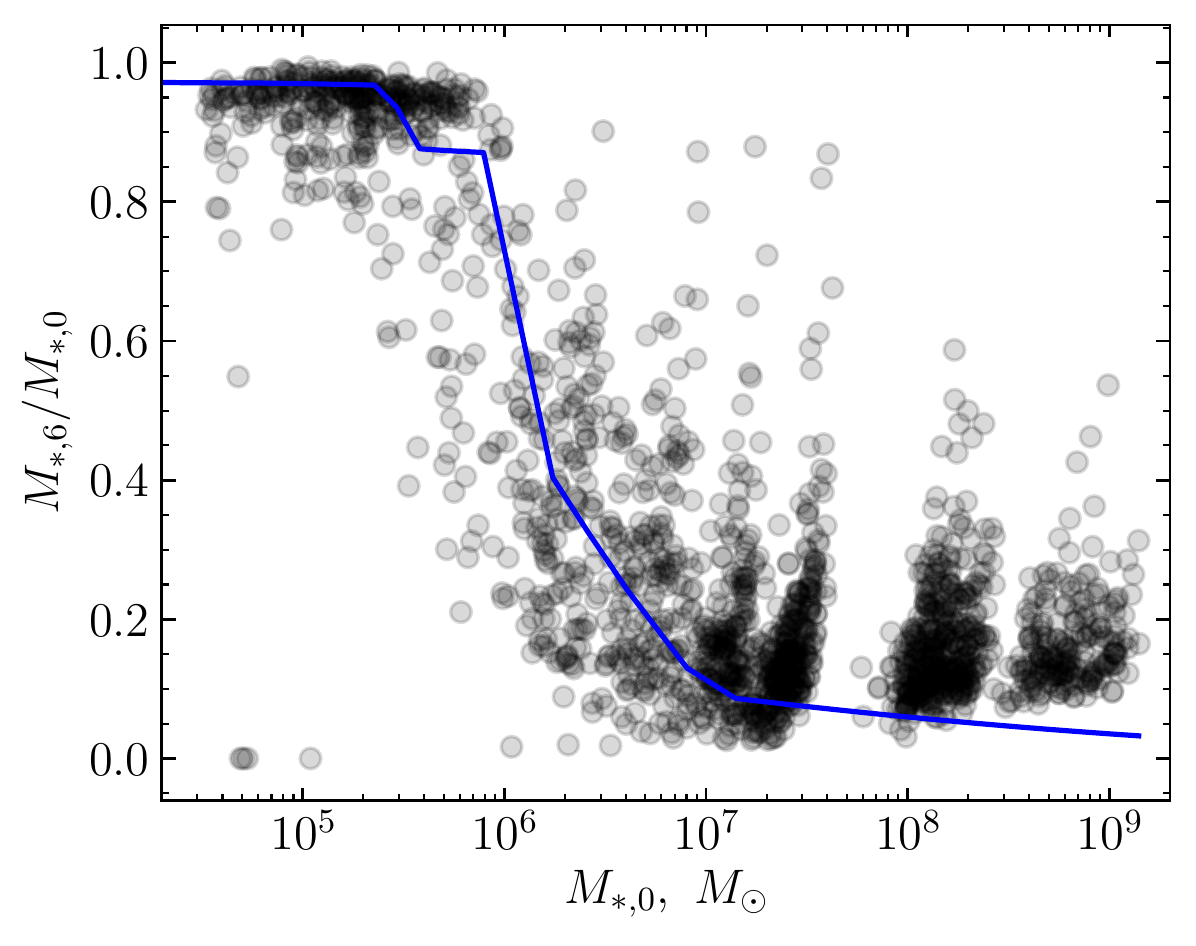}
    \caption{Fraction of stars formed before $z=6$ as a function of present-day stellar mass in average (blue line) and stochastic models (grey circles).}
    \label{fig:SF_before}
\end{figure}

SFR in our model is directly proportional to gas mass, therefore the
total stellar mass (Fig \ref{fig:avg_evolution}, panel c) shows
similar evolutionary stages as the gas mass. Additionally, stellar
mass may grow in fully quenched galaxies due to mergers. Since we do
not deal with processes during which the stellar mass could decrease,
e.g. tidal stripping, stellar mass grows even in the least massive
galaxies. It is clear that stars with low present-day halo and stellar
masses form most of their stellar mass at high redshift; in fact,
galaxies with $M_{*,0} \leq 3\times10^5 \, \msun$ (five lowest-mass
models) have more than half their present-day stellar mass in place by
$z=6$; we show below that the mass increase since then is actually
only due to mergers, rather than in-situ star formation. The most
massive galaxies, conversely, form the majority of their stars after
reionization.

The computed star formation histories can be compared with
observations of dwarf galaxies in the Local Group
\citep{Weisz2014ApJ}. While they exhibit a wide variety of star
formation histories, a general trend of smaller galaxies forming most
of their stars at earlier epochs is evident. In particular, galaxies
with present-day stellar mass $M_{*,0} \leq 10^5 \, \msun$ have, on
average, formed $f_6 \sim 80\%$ of their stars by $z = 6$, although
the spread in this fraction is large, $25\% < f_6 < 90\%$ \citep[see
  Figure 10 of][]{Weisz2014ApJ}. In galaxies with $10^5 \, \msun <
M_{*,0} \leq 10^6 \, \msun$, the average fraction drops to $f_6 \sim
40 \%$. We show our model results in Figure \ref{fig:SF_before}, in
blue line for average models and grey circles for stochastic models
(see Section \ref{sec:sf_stoch} below). The smallest model galaxies,
with $M_{*,0} \simlt 4\times10^5 \, \msun$, have $f_6 \simeq
90-100\%$, with the rest of their stellar mass built up only through
mergers. In more massive galaxies, this fraction drops: in galaxies
with $M_{*,0} > 10^7 \, \msun$, we find $f_6 \leq 10\%$. It is
important to note, however, that this fraction only accounts for star
formation in the main branch of the merger tree that eventually forms
the present-day galaxy. Since the largest galaxies are merging with
many smaller haloes that have $f_6 > 10\%$, the actual fraction of
stars in the present-day galaxy with formation times later than $z=6$
would be larger than $10\%$. This becomes evident when analysing the
star formation histories of the stochastic models (Section
\ref{sec:sf_stoch}). The rapid transition between quenched and
non-quenched populations at $M_{*,0} \simeq 2\times10^6 \, \msun$
suggests that there should be a clear break in mass-weighted ages of
the stellar populations of dwarf galaxies with different stellar
masses.

The gas metallicity evolution shown in Figure \ref{fig:avg_evolution}, panel
d, can be understood qualitatively from gas and stellar mass
evolution. In the smallest galaxies, star formation continues for a
while after the shut-off of smooth gas accretion, enriching the ISM
rapidly because it is not diluted by the infalling low-metallicity IGM
gas. Once the remaining gas is depleted, mainly by evaporation, star
formation ceases and gas metallicity remains constant until the
present day. In the most massive galaxies metallicity starts to
increase faster at $z < 3$ because of enrichment of the IGM material
accreting on to those galaxies, in addition to in-situ enrichment by
stars. It is worth noting that once smooth gas accretion shuts off in
a galaxy, its gas metallicity may increase rapidly to values larger
than those in a more massive galaxy that is still accreting gas from
the IGM (exemplified by the crossing of the violet, brown and pink
lines, the fourth, fifth and sixth most massive galaxy models, at
intermediate redshifts). Therefore high-redshift dwarf galaxy
populations may show an inverted slope in some part of their
mass-metallicity relation.

\subsection{Star formation in stochastic models} \label{sec:sf_stoch}

\begin{figure}
	\includegraphics[width=0.95\columnwidth]{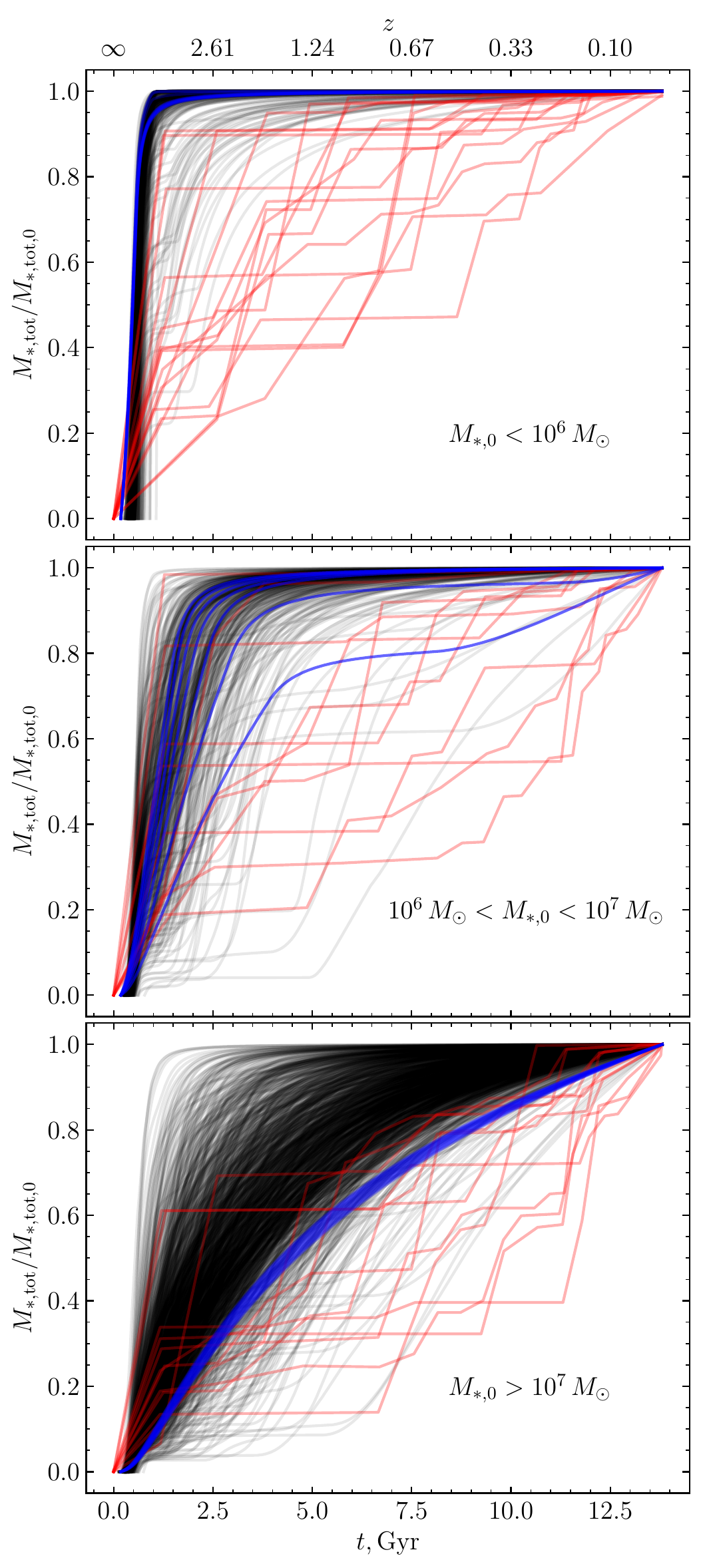}
    \caption{Cumulative star formation histories of model galaxies (grey lines for stochastic models, blue lines for average models) compared with observationally-derived SFHs from \citet[red lines]{Weisz2014ApJ}. Each panel shows a different present-day stellar mass bin, as indicated.}
    \label{fig:fractional_SFH}
\end{figure}

\begin{figure}
	\includegraphics[width=\columnwidth]{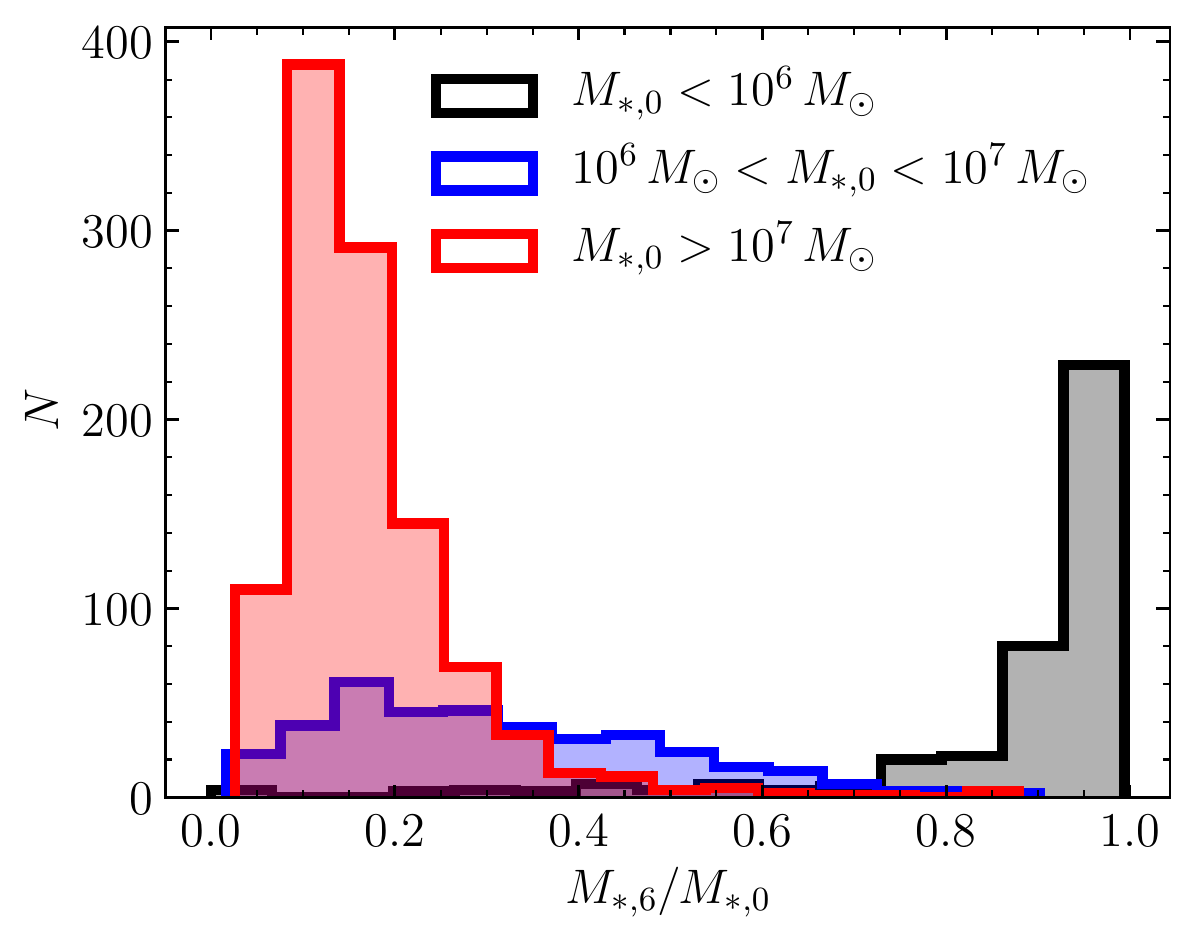}
    \caption{Fractional star formation before $z=6$ in stochastic models of galaxies with different present-day stellar masses: $M_{\rm *,0} < 10^6 \, \msun$ (black), $10^6 \, \msun < M_{\rm *,0} < 10^7 \, \msun$ (blue) and $M_{\rm *,0} > 10^{7} \, \msun$ (red).}
    \label{fig:SF_before_reionization}
\end{figure}

The stochastic models and their star formation histories provide a
picture of how reionization affects galaxies with different mass
assembly histories. Grey circles in figure \ref{fig:SF_before} show
the fraction $f_6$ of stars formed before $z=6$ in the stochastic
models, while figure \ref{fig:fractional_SFH} shows the cumulative
star formation histories of our stochastic model galaxies (grey
lines), average model galaxies (blue lines) and observed Local Group
galaxies from \citep[red lines]{Weisz2014ApJ}. Following
\citet{Weisz2014ApJ}, we separate the galaxies into three bins by
present-day stellar mass.

Two things are immediately clear from these plots. First of all,
stochasticity of mass assembly has a significant effect on the star
formation histories. In small galaxies ($M_{*,0} < 10^6 \, \msun$),
stochastic assembly means that some galaxies are more resilient to
reionisation and therefore can form a significant fraction of their
stars after $z=6$. In massive galaxies ($M_{*,0} > 10^7 \, \msun$),
conversely, stochasticity leads to, on average, older stellar
populations than the average model, because some of these galaxies are
significantly quenched just after reionisation and because the
stochastic models account for dry mergers with small galaxies
containing only very old stars (see penultimate paragraph of Section
\ref{sec:evolution_params}). For intermediate-mass galaxies ($10^6 \,
\msun < M_{*,0} < 10^7 \, \msun$), stochasticity mainly increases the
spread of possible SFHs without changing the mean values.

The second observation to make is that observed galaxies generally
have younger stellar populations than model galaxies, both average and
stochastic ones. This discrepancy is particularly pronounced for the
smallest galaxies: even though stochastic models produce a spread of
the fraction of stars formed before reionisation, with a few outlier
galaxies forming the majority of their stars after $z=6$, the observed
SFHs with very late star formation are not reproduced. Most of our
smallest galaxies are completely quenched by reionisation, in marked
contrast to observations \citep{Weisz2014ApJb}. Similarly, our models
produce more than $60\%$ of stars by $t = 2$~Gyr in galaxies with
$M_{*,0} < 10^6 \, \msun$, but very few observed ones do so. The
differences are somewhat smaller in more massive galaxies.

There may be a few causes of this discrepancy. First of all, our
models do not include any environmental effects. Collisions with
intergalactic gas streams may reignite star formation in even the
smallest galaxies \citep{Wright2019MNRAS}. In fact, spheroidal
galaxies show on average earlier star formation than irregulars
\citep{Weisz2014ApJ}, suggesting that interactions likely played a
significant role in prolonging and/or reigniting their star
formation. Another effect is starbursts caused by gas compression
during major mergers, which we neglect (see Section
\ref{sec:mergers}). Overall, mergers lead to bursts or even prolonged
periods of star formation \citep{Cloet2014MNRAS, Fouquet2017MNRAS}, so
this effect may be important to some extent. On the other hand, the
overall importance of mergers to the SFHs of dwarf galaxies has been
recently questioned \citep{Fitts2018MNRAS}, so it is difficult to
determine the magnitude of this effect. Finally, we assume complete
homogeneity of the Universe, and that reionization happens
simultaneously everywhere. In reality, the growth of HII bubbles was
very patchy \citep{Furlanetto2004ApJ, Furlanetto2005MNRAS,
  Holder2007ApJ}. The effect of patchy growth is cumulative with that
of gas accretion and mergers: in denser regions, reionization occurs
later, gas accretion is faster and mergers more frequent. All three
effects can lead to some galaxies continuing to form stars until much
later times, as seen in detailed numerical simulations
\citep{Katz2019arXiv}.

The trend of star formation histories changing with increasing mass is
qualitatively preserved in our models: in small galaxies, most stars
are formed early, while massive galaxies have prolonged SFHs, with the
switchover happening at present-day halo masses $2\times10^9 \, \msun
\simlt M_{\rm dm,0} \simlt 6\times10^9 \, \msun$, corresponding to
stellar masses $10^6 \, \msun \simlt M_{\rm *,0} \simlt 10^7 \,
\msun$. We show this in Figure \ref{fig:SF_before_reionization}, where
the three histograms present the spread of fractions of stars formed
before reionization in stochastic models of galaxies falling in the
three mass bins described above. It is clear that there is a
transition between early and late star formation modes in this mass
range, with the vast majority of $M_{\rm *,0} < 10^6 \, \msun$
galaxies forming $>80\%$ of their stars before $z=6$, while for
galaxies with $M_{\rm *,0} > 10^{7} \, \msun$, the majority form $<
40\%$ of stars at such early. Galaxies with intermediate stellar
masses represent a transitioning population, with models exhibiting
both significant late star formation and almost complete quenching by
reionization.

\subsection{Build-up of observed correlations}

\begin{figure}
	\includegraphics[width=\columnwidth]{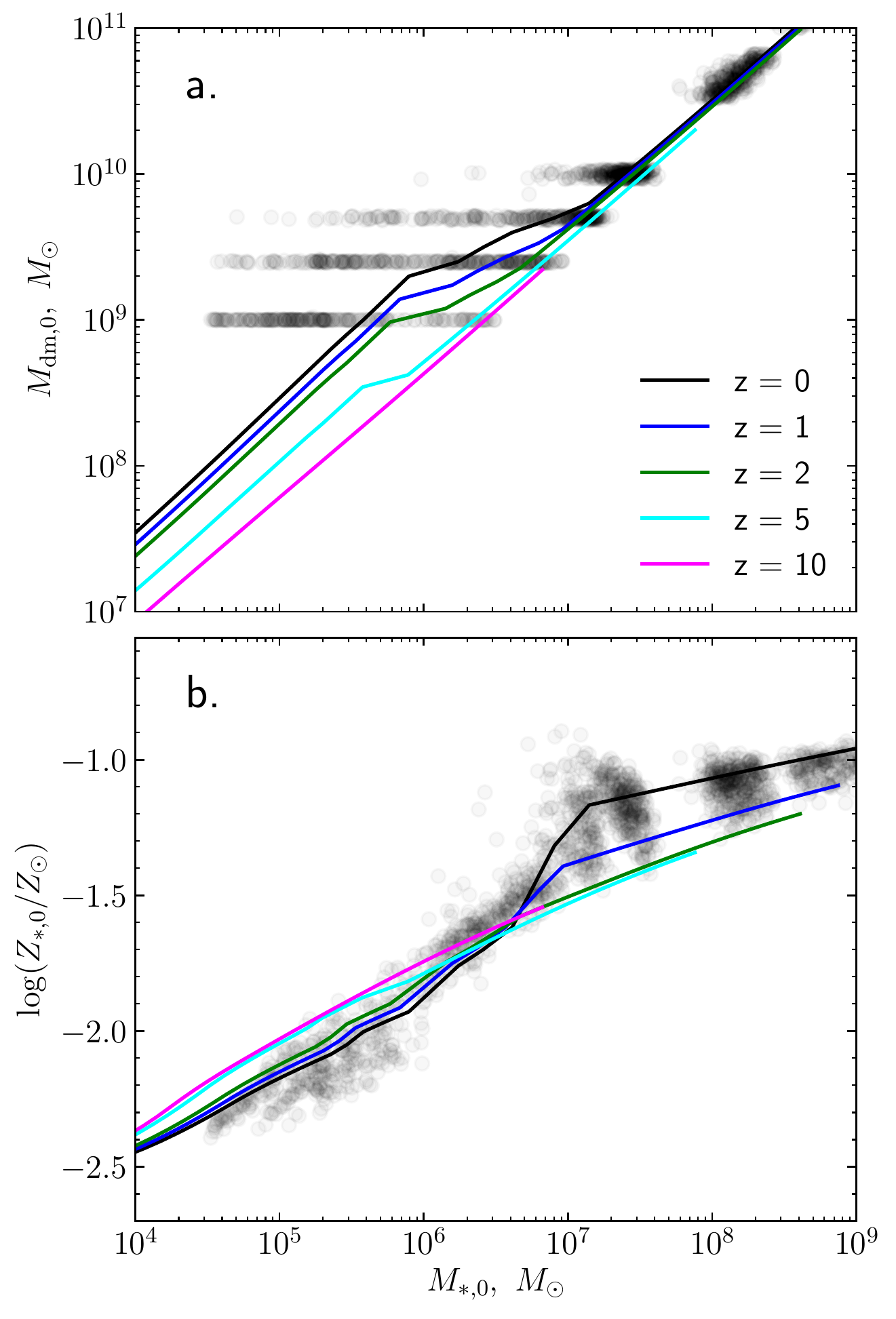}
    \caption{Same relations as in fig. \ref{fig:calib_params} but now differently colored lines denote the average model at different redshifts. The black points represent the tree model with various different mass-assembly histories.}
    \label{fig:relation_evolution_1}
\end{figure}

\begin{figure}
	\includegraphics[width=\columnwidth]{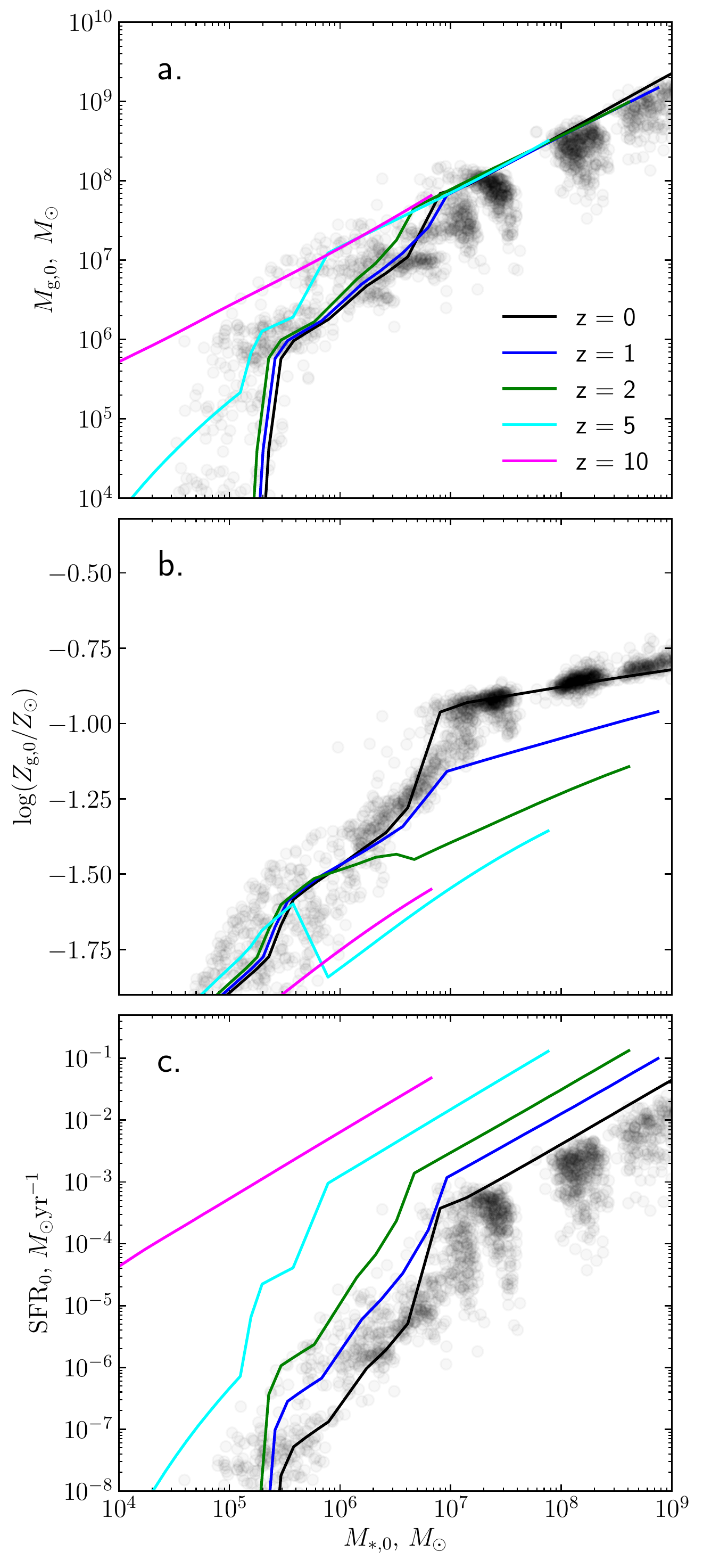}
    \caption{Same relations as in fig. \ref{fig:check_calib} with same markings as in fig. \ref{fig:relation_evolution_1}.}
    \label{fig:relation_evolution_2}
\end{figure}

We now turn to investigating the correlations between present-day
dwarf galaxy parameters, the evolution of these correlations with
redshift, and the spread of individual galaxy parameters around the
average correlations. We use the average models to investigate
relation evolution, and the properties of the 1500 stochastic models
at $z=0$ to investigate individual variations.

Figures \ref{fig:relation_evolution_1} and
\ref{fig:relation_evolution_2} show the same relations as figures
\ref{fig:calib_params} and \ref{fig:check_calib}. However, here we
plot the relations at different cosmological epochs with
different-coloured lines: $z=10$ in magenta, $z=5$ in cyan, $z=2$ in
green, $z=1$ in blue and $z=0$ in black. In addition, semi-transparent
gray circles show the properties of individual galaxies from
stochastic models at $z=0$. These values can be seen as the dispersion
around the average model $z=0$ line due to effects of stochasticity.

The break in the $M_{\rm dm} - M_*$ relation (figure
\ref{fig:relation_evolution_1}, panel a) in the average model is
evident as early as $z = 5$ and persists until the present day, with
the only change being the halo mass at which it occurs. The origin of
the break, as discussed in subsection \ref{sec:average_evolution}, is
the effect of reionisation clearing gas out of the smallest galaxies,
shutting off their star formation. The precise effect on each galaxy
around the break mass depends sensitively on its mass assembly
history, as evidenced by the stochastic models. These show not a clear
break, but a significant spread of stellar masses for each final halo
mass around the break. Some galaxies assemble their halo masses early
on and can continue forming stars until low redshift, while others
spend a significant amount of time quenched by reionisation, sometimes
reigniting at late times (see Paper I). A similar spread is present at
higher redshifts as well (see section \ref{sec:sf_stoch} above).

All other relations show noticeable breaks at stellar masses $M_*
\simeq 10^7 \, \msun$, both in the average and the stochastic
models. It is worth noting, however, that those breaks appear very
differently at higher redshift. For example, the $M_* - Z_*$
relationship shows a break only at $z \leq 1$, while the $M_* - M_{\rm
  g}$ and $M_* - $SFR relationships show a clear break as early as
$z=5$, although the stellar mass at which the break occurs increases
with time. The $M_* - Z_{\rm g}$ relationship is even more different,
with an inverse break at $2\leq z\leq5$, no break at $z = 10$ and a
positive break at $z\leq1$. All of these breaks also arise due to
reionization, but the timescale for them to appear is different. Gas
mass and the closely related SFR decrease very rapidly in the smallest
galaxies. Stellar metallicity, on the other hand, does not decrease,
but merely stops increasing in those galaxies, while it continues to
grow in the larger ones, as evident from figure
\ref{fig:relation_evolution_1}, panel b, when looking at the average
model curves. Gas metallicity increases rapidly in quenched galaxies,
creating a negative slope, until the stars in more massive galaxies
have enough time to `catch up' and enrich the gas further. This
variation in the evolution of observable relations suggests that
high-redshift dwarf galaxies may exhibit very different correlations
between parameters than local ones. In particular, differences due to
reionization only manifest strongly at $z<2$, and become more
pronounced with time, so local galaxies may provide better evidence of
reionization effects than high-redshift ones.

The stochastic models reveal how the variation of mass assembly
affects the relations. Both metallicity relations exhibit a spread
around the $z=0$ value of the average model. The spread of stellar
metallicities is $\pm \sim0.25$~dex around the break mass, and smaller
for both more and less massive galaxies. Gas metallicity has a large
spread of $\pm \sim0.2$~dex at low galaxy masses, but practically zero
spread for galaxies above the break, which all experience rather
similar growth of stellar mass, and hence enrichment of the ISM.

The gas masses in stochastic models are systematically smaller than in
the average model. Correspondingly, the SFRs are lower as well. The
reason for this is that stochastic models spend some time at lower
halo masses than corresponding average models, as well as some time at
higher masses. However, being at a lower halo mass can bring a given
stochastic model under the threshold for gas evaporation, while being
at a higher mass does not give a corresponding advantage to gas
accretion. As a result, stochastic models tend to lose more gas to
evaporation than average models. This is evident to some extent in the
$M_{\rm dm} - M_*$ relation, where stochastic models have a bigger
spread toward lower stellar masses at a given halo mass than toward
higher stellar masses. This effect slightly reduces the alignment of
our model results to the observational data (figure
\ref{fig:check_calib}, panel c).

Comparing the average models at high redshift to both average and
stochastic models at low redshift reveals that galaxies at high
redshift were significantly different from modern-day ones, especially
in their gas properties and SFRs. This evolution of gas mass, gas
metallicity and SFR at a given $M_*$ or $M_{\rm dm}$ occurs even
though these galaxies are all isolated. The extreme evolution in SFRs
(SFR decreases by almost four orders of magnitude from $z=10$ for
galaxies with $M_* = 10^6 \, \msun$ and two orders of magnitude for
$M_* = 10^7 \, \msun$) occurs both because of gas depletion and
because of passive evolution of galaxy sizes: modern galaxies are less
compact, therefore have lower gas surface densities and lower
molecular gas fractions, which lead to lower SFRs even for identical
gas masses.

\section{Discussion}
\label{sec:discussion}

\subsection{Trends of dwarf galaxy properties}

Qualitatively, the evolution of the $M_* -$ SFR relation that we find
in our models agrees with the observed evolution \citep{Tasca2015A&A,
  Santini2017ApJ}, although the latter is derived for more massive
galaxies. In particular, the specific SFR (SFR$/M_*$) increases with
redshift, and galaxies have similar specific SFRs across the mass
range at a given redshift. Quantitatively, the specific SFR in the
most massive galaxies in our models is lower by a factor of a few than
the average observational result, both at $z=0$ \citep{Peng2010ApJ}
and at higher redshift \citep{Tasca2015A&A, Santini2017ApJ}. The SFRs
observed in low-mass galaxies agree very closely with our model
results, however (figure \ref{fig:check_calib}, panel c). The reason
for this is that our models are calibrated using data of isolated
dwarf galaxies, which comprise only a small fraction of typically
observed galaxies. They tend to have lower SFRs than average, since
they are not affected by their environment, which may cause shocks and
lead to starbursts.

The predicted trend of the mass-metallicity relation, if extended to
higher stellar masses, would also give slightly lower metallicities
than observed \citep[$Z \simeq Z_\odot$ at $M_* = 10^9 \, \msun$,
  cf.][]{Sanchez2017MNRAS}. It should be noted, however, that
metallicities decrease steeply at $M_* < 10^9 \, \msun$, and the data
available for these smaller galaxies generally agrees with our results
(figure \ref{fig:calib_params}, panel b). Matching the trend obtained
in our models to that observed for more massive galaxies is not
possible without a more detailed treatment of enrichment of the ISM,
the escape of metals into the IGM, and effects such as the galactic
fountain.

The $M_* - M_{\rm g}$ relation agrees very well with that observed in
small galaxies, but observations show a break at $M_* \simeq
10^9-10^{10} \, \msun$, with more massive galaxies having a shallower
slope \citep{Lelli2016AJ,McGaugh2017ApJ}. Therefore we expect that our
model would overpredict the gas masses of more massive galaxies. This
is expected, since AGN outflows, which are not included in our model,
should become significant to galaxy evolution at masses $M_* \simgt
10^{10}\, \msun$ \citep{Martin2018ApJ, Zubovas2018MNRAS}.

\subsection{Stochasticity of dwarf galaxy evolution}

There is a large volume of work exploring the importance to dwarf
galaxy evolution of various stochastic effects, such as star formation
\citep{Gerola1980ApJ, Matteucci1983A&A, Orban2008ApJ, Weisz2008ApJ,
  Mineikis2010BaltA, Weisz2012ApJ, Applebaum2020MNRAS}, reionization
\citep{Bose2018ApJ, Katz2019arXiv}, mergers \citep{Laporte2015MNRAS,
  Benitez2015MNRAS, Maccio2017MNRAS, Angles2017MNRASb} or other
environmental interactions \citep{Penarrubia2012ApJ}. In addition,
stochastic processes may affect the appearance and inferred properties
of dwarf galaxies, e.g. their star formation rates
\citep{daSilva2014MNRAS}, H$\alpha$ luminosities
\citep{Fumagalli2011ApJ} or their contribution to reionization
\citep{Forero2013MNRAS}. Our work contributes to this field by looking
at stochasticity from another angle: the differences in halo assembly
histories of individual galaxies, which is related to, but somewhat
broader in scope than, the stochasticity of merger events.

The differences between properties of galaxies in average and
stochastic models show that stochasticity of mass assembly is an
important element of dwarf galaxy evolution, in general agreement with
semi-analytical models based on N-body simulation results
\citep{Yaryura2016MNRAS}. Stochasticity induces a spread in the
observable relations around the mean value. Additionally, for galaxies
with $M_* > 10^7 \, \msun$, our stochastic models predict
significantly lower gas masses and SFRs for a given value of $M_*$
than average models; for galaxies with $M_* < 10^7 \, \msun$,
stochastic models predict higher average halo masses and somewhat
higher gas masses and SFRs.

The main reason for these systematic differences is the different
responses of halos to reionization. The more massive a given halo is
at $z = 6$, the less it is affected by reionization, the more gas it
retains and the more stars it can form. Halos with present-day masses
$M_{\rm dm,0} > 10^{10} \, \msun$, typically hosting galaxies with
$M_{*,0} > 10^7 \, \msun$, have large enough masses at $z = 6$ in the
average model that they are affected very little by
reionization. Halos in stochastic models have both larger and smaller
masses than those in the average model, but those with higher masses
evolve very similarly to the average ones; conversely, those with
lower masses are affected more significantly by reionization. Later
on, these halos merge with smaller galaxies that retain much less gas,
leading to lower gas masses and SFRs at lower redshift. For smaller
halos, variations of mass assembly history become more
significant. Halos that build up their mass early are less affected by
reionization, retain their gas and form stars efficiently, generally
following the trends in $M_* - M_{\rm dm}$ and other relations set by
larger galaxies. However, halos that assemble later are strongly
affected, leading to earlier truncation of star formation and lower
present-day stellar mass. Some of these galaxies can be re-ignited at
later times if mergers bring their masses above the evaporation
threshold (see Paper I), therefore a fraction of these galaxies have
higher gas masses and SFRs than predicted by the average model.

The spread in halo mass, gas mass, stellar and gas metallicity, and
star formation rate at a given stellar mass in our stochastic models
(Figures \ref{fig:relation_evolution_1} and
\ref{fig:relation_evolution_2}) is similar to that in real galaxies
\citet[Figures \ref{fig:calib_params} and \ref{fig:check_calib}; also
  see][]{Read2017MNRAS, m_s_Z_s_relation, cuspvscore, Kirby2017,
  m_s_Z_g_relation, McGaugh2017ApJ}. It appears that stochastic
differences of halo mass assembly can account for most of the spread
of observed gas masses and star formation rates in modern-day dwarf
galaxies without invoking any other physical processes.

We deliberately chose to include only this aspect of stochasticity
while neglecting other potential sources. In particular, we fixed the
redshift of the end of reionization for all simulations, the star
formation prescription is purely deterministic and there are no
environmental interactions included. Taken together, these sources of
stochasticity may account for the whole spread of dwarf galaxy
properties. It would be interesting to see how these processes
interact with each other and what their relative contributions to
total variation of dwarf galaxy parameters are, but such an
investigation is beyond the scope of this paper. Nevertheless, our
simulations show that stochasticity of mass assembly should be kept in
mind when interpreting observed dwarf galaxy properties and following
their evolution with semi-analytic models.

Acknowledging and understanding the importance of stochasticity will
also help make better predictions regarding observable dwarf galaxy
trends. Some of the parameters in the stochastic models, e.g. gas and
stellar metallicity, follow the average models quite closely, while
others, e.g. gas mass and SFR, are spread out rather widely. The
latter parameters are therefore poor choices when trying to constrain
the position of the break in dwarf galaxy properties at $M_* \sim 10^7
\, \msun$, since identification of the break would require a large
number of data points with small errors. The mass at which the break
occurs depends on the process of reionization \citep{Bose2018ApJ,
  Katz2019arXiv} and can help us infer its properties, therefore it is
important to know whether the observed breaks are due to individual
stochastic variations or global trends.

\subsection{Starbursts and star formation quenching}

Our stochastic models do not show any starburst dwarf galaxies with
specific star formation rates SFR$/M_* > 1$~Gyr$^{-1}$. To some
extent, this is an artifact of our model, where we do not account for
possible shocks during galaxy mergers increasing the efficiency of
star formation, not do we account for the triggering of star formation
by supernovae or AGN effects. The observed starburst fraction does not
depend strongly on galaxy mass \citep{Bergvall2016A&A}, so we may
expect a few percent of dwarf galaxies to also be in starburst mode at
the present time. The overall influence of starbursts on integrated
galaxy properties appears to be small \citep{Bergvall2016A&A},
therefore the lack of starbursts in our model should not affect the
overall results very much.

Some of the galaxies in the stochastic models have very low gas masses
and, as an extention, very low SFRs. Their specific SFRs can fall as
low as $10^{-3}$~Gyr$^{-1}$. Such galaxies may correspond to the
observed dwarf spheroidal (dSph) galaxies, which are generally
quenched. The quenching in our models is, however, not a consequence
of a starburst or some other internal process, but rather of the mass
assembly history preventing gas accretion on to them for significant
periods of time. Virtually all galaxies with $M_* < 10^7 \, \msun$
fall into this group and a small fraction of more massive ones also
do. As more observational data is collected, population analysis of
dwarf galaxy specific star formation rates may provide insight into
the importance of internal versus external quenching processes. If
dSph and other types of quenched dwarf galaxies form a significant
fraction of $M_* > 10^7 \, \msun$ galaxies at $z=0$, it would mean
that starbursts and other internal processes are important quenching
channels, since the population would not be explained purely by
different mass assembly histories and the effect of the ionizing
background radiation. In the future, we plan to extend our model to
include better treatment of starbursts, AGN feedback and galaxy-galaxy
interactions to account for these effects.

\subsection{Galaxy properties at very low stellar masses} \label{sec:lowmass}

Observational data of dwarf galaxies below $M_* \sim 10^7 \, \msun$ is
scarce \citep[e.g.,][]{m_s_Z_g_relation, cuspvscore, Kirby2017,
  m_s_m_dm_relation, McGaugh2017ApJ}, therefore checking the existence
of the break predicted by our model is difficult. The few data points
at lower stellar masses generally seem to agree with our results, but
it will be very interesting to get more information about these
smallest dwarf galaxies. Observations of high-redshift galaxies and
their properties also have only limited overlap with the mass range of
dwarf galaxies \citep[e.g.,][]{Santini2017ApJ}. Therefore it will be
very important to test the model with larger data sets extending down
to lower masses and higher redshift, such as could be provided by the
upcoming Euclid space observatory \citep{Laureijs2011arXiv}. Most
importantly, they will help better understand the transition region
between dwarf galaxies, dominated by stellar feedback, and
AGN-dominated massive galaxies. In addition, our model should help
better understand the relative importance of internal (e.g. supernova
feedback) and external (e.g. intergalactic radiation field) processes
to dwarf galaxy evolution.

The relationship between gas metallicity and stellar mass is
particularly interesting and may be the most illuminating when it
comes to high-redshift galaxy properties. The reversal of its trend,
with metallicity decreasing with increasing stellar mass, over a
certain mass range for a range of redshifts, should help us better
understand the enrichment of the ISM and IGM by stellar processes (see
sections \ref{sec:chem_evolution} and \ref{sec:igm}).

\section{Conclusions}\label{sec:conclusions}

In this paper, we presented results of a semi-analytical model
following the evolution of isolated dwarf galaxies from early Universe
until the present day. Our model uses a few free parameters, but when
they are calibrated using the halo mass - stellar mass and stellar
mass-metallicity relations, the model fits other observed properties
of the local dwarf galaxy population remarkably well. The relations
are generally affected very strongly by reionization, which suppresses
gas infall in the smallest galaxies, even evaporating some of
them. Different relations acquire their characteristic shapes at
different times. At the present day, most of them exhibit a break at
$M_* = 10^7 \, \msun$, with smaller galaxies being significantly
affected by reionization, and larger ones being affected only
indirectly, by changes to the small galaxies that merged with them
over the Hubble time. The position of the break depends on the
properties of reionization, so in principle could be used to test
these properties, once larger samples of isolated dwarf galaxy
properties become available.

We also find that the stochasticity of mass assembly histories has a
strong effect on dwarf galaxy gas masses and SFRs. Models not
accounting for this effect may overpredict the gas mass and SFR for a
galaxy with a given dark matter halo and stellar mass, leading to
incorrect determination of dwarf galaxy properties or incorrect
calibration of model parameters.

Our results suggest that future observations, that will reveal large
samples of low-mass high-redshift galaxies, will be instrumental in
understanding the evolution of these galaxy building blocks.

\section*{Acknowledgements}

We thank Vladas Vansevi\v{c}ius for valuable comments on the draft
version of this paper and the anonymous referee for suggestions that
helped improve the clarity of the argument. This research was funded
by a grant (No. LAT-09/2016) from the Research Council of Lithuania.





\bsp	
\label{lastpage}
\end{document}